# Blip-Up Blip-Down Circular EPI (BUDA-cEPI) for Distortion-Free dMRI with Rapid Unrolled Deep Learning Reconstruction

*Uten Yarach[1#], Itthi Chatnuntawech[2#], Congyu Liao[3,4], Surat Teerapittayanon[2], Siddharth Srinivasan Iyer[5], Tae Hyung Kim[6], Justin Haldar[7], Jaejin Cho[8], Berkin Bilgic[8,9], Yuxin Hu[10], Brian Hargreaves[3,4,10], and Kawin Setsompop[3,4*]*

*1 Radiologic Technology Department, Associated Medical Sciences, Chiang Mai University, Chiang Mai, Thailand*

*2 National Nanotechnology Center, National Science and Technology Development Agency, Pathum Thani, Thailand*

*3 Department of Radiology, Stanford University, Stanford, CA, USA*

*4 Department of Electrical Engineering, Stanford University, Stanford, CA, USA*

*5 Department of Electrical Engineering and Computer Science, Massachusetts Institute of Technology, Cambridge, MA, USA*

*6 Department of Computer Engineering, Hongik University, Seoul, South Korea*

*7 Signal and Image Processing Institute, Ming Hsieh Department of Electrical and Computer Engineering, University of Southern California, Los Angeles, CA, USA*

*8 Athinoula A. Martinos Center for Biomedical Imaging, Massachusetts General Hospital, Charlestown, MA, USA*

*9 Department of Radiology, Harvard Medical School, Boston, MA, USA*

*10 Department of Bioengineering, Stanford University, Stanford, California, USA.*

*[#] These authors have contributed equally to this work and each should be regarded as first author*

*\* Corresponding author*

*Kawin Setsompop, PhD, kawins@stanford.edu, Room 365, Packard Electrical Engineering Building, 350 Jane Stanford Way, Stanford, CA, 94305*

**Short running head:** BUDA-cEPI RUN-UP Reconstruction for Distortion-Free dMRI

Word Count:    Abstract: 247

Manuscript:    4725

Number:    Figures: 8

Tables: 3

References: 71

ABSTRACT

Purpose: We implemented the blip-up, blip-down circular echo planar imaging (BUDA-cEPI) sequence with readout and phase partial Fourier to reduced off-resonance effect and T2* blurring. BUDA-cEPI reconstruction with S-based low-rank modeling of local k-space neighborhoods (S-LORAKS) is shown to be effective at reconstructing the highly under-sampled BUDA-cEPI data, but it is computationally intensive. Thus, we developed an ML-based reconstruction technique termed "BUDA-cEPI RUN-UP" to enable fast reconstruction.

Methods: BUDA-cEPI RUN-UP – a model-based framework that incorporates off-resonance and eddy current effects was unrolled through an artificial neural network with only six gradient updates. The unrolled network alternates between data consistency (i.e., forward BUDA-cEPI and its adjoint) and regularization steps where U-Net plays a role as the regularizer. To handle the partial Fourier effect, the virtual coil concept was also incorporated into the reconstruction to effectively take advantage of the smooth phase prior, and trained to predict the ground-truth images obtained by BUDA-cEPI with S-LORAKS.

Results: BUDA-cEPI with S-LORAKS reconstruction enabled the management of off-resonance, partial Fourier, and residual aliasing artifacts. However, the reconstruction time is approximately 225 seconds per slice, which may not be practical in a clinical setting. In contrast, the proposed BUDA-cEPI RUN-UP yielded similar results to BUDA-cEPI with S-LORAKS, with less than a 5% normalized root mean square error detected, while the reconstruction time is approximately 3 seconds.

Conclusion: BUDA-cEPI RUN-UP was shown to reduce the reconstruction time by ~88x when compared to the state-of-the-art technique, while preserving imaging details as demonstrated through DTI application.

# 1. INTRODUCTION

Diffusion represents alterations in the random movement of water molecules in tissues, revealing their microarchitecture, and microstructural abnormalities in many neurological conditions. Diffusion magnetic resonance imaging (dMRI) provides useful information, increasing the sensitivity of MRI as a diagnostic tool, narrowing the differential diagnosis, and providing prognostic information for treatment planning[1,2]. Single-shot Echo-Planar Imaging (ssEPI) is widely used for clinical dMRI since it is one of the fastest imaging techniques[3]. However, for high resolution dMRI, ssEPI is often compromised by off-resonance and eddy current effects which cause geometric distortion and T2* decay which causes blurring, due to the lengthy echo spacing (ESP) and echo train length (ETL).

Numerous techniques have been developed to mitigate the aforementioned issues[4-13]. Post-processing techniques that require gradient echo or EPI based field maps attempt to correct geometric distortion through image-domain interpolation[8,9] However, due to finite data sampling and image discretization, interpolative resampling invariably causes image blurring and spatial resolution loss. Moreover, because these non-idealities do not manifest independently during the data acquisition, traditional post-processing techniques that independently manage them can leave (potentially subtle) residual errors or degradations in the resulting images. Model-based reconstructions[11-13], that consider off-resonance, odd-even phase shift, gradient nonlinearity, and/or ramp sampling in signal forward model and reconstruct images via iterative least-squares solver, have been shown to effectively mitigate some residual artifacts that occur in standard post-processing corrections.

For high resolution dMRI where ESP is extended, model-based reconstruction alone may be insufficient to mitigate distortions in ssEPI. To overcome this issue, interleaved multi-shot EPI (msEPI) acquisition can be used in conjunction with model-based reconstruction, where the effective ESP is shortened by a factor equal to that of the shot number used, at the expense of prolonged scan time[14-16]. Modified rapid 'two-shot' EPI acquisitions such as blip-up, blip-down EPI acquisition (BUDA) and related techniques have also been developed to improve on this and shown to enable high-fidelity, high-resolution distortion-free dMRI[17-22] In BUDA, the off-resonance maps are often estimated from the individual blip-up and -down images which include both B0 field inhomogeneity and eddy current effect. With the incorporation of these maps, two-shot data can be jointly reconstructed through a model-based framework with low rank matrix modeling constrained algorithms[23]. These algorithms enable handling of shot-to-shot background phase variations and partial Fourier (pF) sampling effect, thereby providing high-fidelity dMRI without the need of additional calibration data[24-28]. However, the associated long reconstruction time[29] has limited the use of these algorithms in clinical applications.

In the past few years, two types of deep learning-based strategies have been adopted to reduce MRI reconstruction time: data-driven[30-35] and model-driven approaches[36-43]. The data-driven approach typically trains a standard neural network model, such as a multi-layer perceptron (MLP) and a convolutional neural network (CNN), using a collection of input-output pairs to approximate the unknown underlying input-output relationship. In the context of accelerated dMRI, the input and output to the model can be chosen to be undersampled k-space and the ground truth images for simplicity, respectively. While such an approach has demonstrated promising results with much shorter reconstruction time compared to the conventional iterative reconstructions, it lacks theoretical explanation on the relationship between the network topology and performance. Furthermore, successful applications of the approach typically require a large amount of data, which could be clinically prohibitive. For the model-driven approach, an optimization problem - that relates the input data to the output data is formulated based on MR physics, and an optimization algorithm to solve the formulated problem is then selected and unrolled, resulting in a deep learning model with the architecture that is tailored to the current application. By explicitly incorporating domain-specific knowledge into a deep learning model this way, the model-based approach can become more robust to scan- and/or subject-specific factors and relies less on massive datasets for training. Recently, RUN-UP[38], a model-driven deep learning approach, was proposed for multi-shot dMRI reconstruction to speed up the reconstruction. Specifically, fast iterative shrinkage-thresholding algorithm (FISTA)[44] was unrolled for a fixed number of iterations. The unrolled network alternates between data consistency (i.e., forward SENSE and its adjoint) and regularization steps similar to the conventional algorithm, in which U-Net plays a role as a regularizer.

In this work, we propose to use the BUDA circular-EPI (BUDA-cEPI) sequence[45] to produce distortion-free high resolution dMRI. The echo time (TE) and ETL were minimized through partial Fourier (pF) acquisition that is applied on both the readout (RO) and phase encoding (PE) directions. The k-space centers of the blip-up and -down shots were acquired using a constant ESP to enable reconstruction of the individual blip-up and -down low-resolution images at a constant distortion level. These low-resolution images can then be used to accurately estimate off-resonance and eddy current effects which are incorporated into the joint reconstruction. To enable fast reconstruction, we extend the RUN-UP model[38] by incorporating the BUDA-cEPI operator and virtual coil concept. The main contributions of this work are as follow:

- Propose a generalized cEPI signal model that includes off-resonance effect under variable ESP.
- Propose a rapid unrolled model-based framework for reconstructing accelerated BUDA-cEPI, which is approximately 88x faster that the state-of-the-art technique

## 2. THEORY

### 2.1 BUDA Circular EPI (BUDA-cEPI) Sequence

The sequence-diagram of BUDA-cEPI is illustrated in Fig. 1(A), where two interleaved EPI shots sample complementary subsets of k-space using opposing phase-encoding directions to create opposing distortions. As shown in Fig.1(B), cEPI only acquires approximately 30% of the k-space, using both RO- and PE-pF and ramp-sampling, which significantly reduces ETL and TE. The pF is designed to sample complementary k-space regions across the blip-up and -down shots to enable effective recovery of missing pF data when a joint reconstruction is performed across shots. The regions near the k-space center are acquired using a constant ESP, allowing for reconstruction of blip-up and -down low-resolution images at a constant distortion level that can later be used for off-resonance map estimation. The low-resolution images are in 2 mm. resolution which is high enough to be able to represent the expected shot-to-shot background phase. Both shots are acquired in an interleaved fashion. The sequence programming is implemented using GE EPIC with KS Foundation (https://ksfoundationepic.org/).

## 2.2 Circular EPI Signal Model

Let TE, Δt, and T denote the echo-, dwell-, and echo spacing - times, respectively. Neglecting the $T_2^*$ effect, the signal measured from the object field-of-view ($\Omega_{xy}$) during readout m ∈ {1, ..., M} of phase-encoding line n ∈ {1, ..., N} can be modeled as:

$$g[m, n, c] = \int\int_{\Omega_{xy}} s_c(x,y) f(x,y) e^{-j(\Delta\omega_0(x,y)(\tau[m,n]))} e^{-j(k_x[m]x + k_y[n]x)} + \varepsilon[m, n, c] \quad (1)$$

where f is the target image. $k_x$ and $k_y$ are the k-space coordinates in the readout/frequency and phase-encoded dimensions, respectively. $s_c$ is the sensitivity profile for coil c ∈ {1, ..., C}. $\Delta\omega_0(x, y)$ is the off-resonance caused by magnetic field inhomogeneity at location (x, y). $\tau[m, n] = TE + \left(m - \frac{M-1}{2}\right)\Delta t + \left(n - \frac{N-1}{2}\right)T_n$, denoting the sampling time where $T_n$ is the variable ESP at phase encoding n. ε is Gaussian noise. Eq. 1 accommodates time reversal, odd-even echo shift due to gradient time delay, ramp sampling, parallel imaging, and partial Fourier acceleration.

## 2.3 Discrete Circular EPI Signal Model

As cEPI uses high readout bandwidths, Δt ≪ T and off-resonance primarily manifests along its phase encoded direction. Thus, Δt ≈ 0 can be assumed. Letting $u(x, y) = f(x, y) e^{-j\Delta\omega_0(x,y)TE}$, followed by data discretizing[46], Eq. (1) becomes

$$g[m, n, c] = \sum_{p=1}^{P} \sum_{q=1}^{Q} s_c[p,q] u[p,q] e^{-j(\Delta\omega_0[p,q][n-\frac{N-1}{2}]T_n)} e^{-j(k_x[m]p + k_y[n]q))} + \varepsilon[m, n, c] \quad (2)$$

p and q are the pixel indices. u is the underlying image. When the time at any phase encoding line is assumed to be constant, time segmentation[47] can be applied. Defining $W_n = \text{diag}\{e^{-j(\Delta\omega_0[p,q][n-\frac{N-1}{2}]T_n)}\}$ and $\mathbb{S} = [\text{diag}\{s_1\}, \cdots, \text{diag}\{s_C\}]^T$, Eq. (2) can be written as

$$G = \left(I \otimes \sum_{n=1}^{N} FW_n\right) \mathbb{S}u + \mathcal{E} = Au + \mathcal{E}, \quad (3)$$

where $I$ is the identity matrix, ⊗ is the Kronecker product and $N$ is the number of time segmentation (i.e., total phase encoding lines). F is Fourier transform.

## 2.4 BUDA-cEPI Reconstruction with S-LORAKS

Low-rank modeling of local k-space neighborhoods (LORAKS)[48] is a constrained MRI framework that enables accurate image reconstruction from sparsely and unconventionally sampled k-space data. It has demonstrated that k-space data for MR images with limited spatial support or slowly varying image phase can be mapped into structured low-rank matrices. Moreover, low-rank matrix regularization techniques can be applied to these matrices to produce high-quality reconstructions. In this work, the partial and parallel BUDA-cEPI acquisition were modeled using two matrices: $A_\uparrow$ and $A_\downarrow$ for the blip-up and blip-down acquisitions, respectively. To reconstruct the underlying images $u_\uparrow$ and $u_\downarrow$ with S-LORAKS constraint, we minimize the following objective function $\min_{u_\uparrow, u_\downarrow} \frac{1}{2} \left\| \begin{bmatrix} A_\uparrow & 0 \\ 0 & A_\downarrow \end{bmatrix} \begin{bmatrix} u_\uparrow \\ u_\downarrow \end{bmatrix} - \begin{bmatrix} G_\uparrow \\ G_\downarrow \end{bmatrix} \right\|_2^2 +$ $\lambda J_r(P_{s\uparrow\downarrow}(u_\uparrow, u_\downarrow))$. (4)

$P_{s\uparrow\downarrow}(\cdot)$ is the operator that constructs the high-dimensional structured LORAKS matrix (i.e., S-matrix)[48] of $u_\uparrow$ and $u_\downarrow$. The regularization term $J_r(\cdot)$ is a nonconvex regularization penalty imposing rank-r approximation of the corresponding input matrix, defined as

$$J_r(\mathbf{X}) = \sum_{k>r} \sigma_k^2 = \min_{rank(\mathbf{Y}) \leq r} ||\mathbf{X} - \mathbf{Y}||_F^2$$

λ is a user-selected regularization parameter used to adjust the strength of the regularization penalty applied to the S-matrix. r is a user-selected rank estimates for the S-matrix. $J_r(\cdot)$ is a nonconvex regularization that encourages its matrix argument to have rank less than or equal to r.

## 2.5 RUN-UP: The Unrolled Network with Deep Priors

RUN-UP[38] was introduced for multi-short DWI, CNN regularization was implemented which is implied to utilize the correlations between images from different shots as follows,

$$\min_{\{u_1,\ldots,u_{N_s}\}} \frac{1}{2} \sum_{s=1}^{N_s} ||A_s u_s - G_s||_2^2 + R(u_1, \ldots, u_{N_s}), \quad (5)$$

where $u_1, \ldots, u_{N_s}$ are the images of $N_s$ different shots to be reconstructed, $A_s$ is the encoding operator for the $s^{th}$ shot, which is a combination of the sampling operator, Fourier transform, and sensitivity encoding operator, $G_s$ is the acquired multi-coil data of the $s^{th}$ shot, and $R(\cdot)$ is a regularization term that is modeled using U-Nets which is trained to predict the ground truth obtained by magnitude-based spatial-angular locally low-rank regularization (SPA-LLR)[49]. In particular, the multi shots images are updated using the following equations.

$$u_{1,t} = u_{s,t-1} - \tau(A_s^H A_s u_{s,t-1} - A_s^H G_s)$$
$$\vdots \quad (6)$$
$$u_{N_s,t} = u_{N_s,t-1} - \tau(A_{N_s}^H A_{N_s} u_{N_s,t-1} - A_{N_s}^H G_{N_s})$$

$$\{u_{1,t+1}, \ldots, u_{N_s,t+1}\} = R(u_{1,t}, \ldots, u_{N_s,t}) \quad (7)$$

$A^H$ is the adjoint of $A$. $\tau$ is the step size. When $t$ is an odd number, $R(\cdot)$ takes k-space data as the input ($F\{u_{\{1,t\}}, \ldots, u_{\{N_s,t\}}\}$). When $t$ is an even number, $R(\cdot)$ takes image data as the input ($u_{\{1,t+even\}}, \ldots, u_{\{N_s,t+even\}}$). This implementation is called KI-Net.

## 3. METHODS

### 3.1 Data Acquisitions

In-vivo experiments were performed on a 3T GE Premier with a 48-channel receiver head coil (SVD-compressed to 12-channel[50]). Ten healthy volunteers were scanned with informed consent according to an IRB protocol. Three protocols shown in Table 1 were performed. *Data-I, Data-II, and Data-III* were used for cEPI's simulation, training, and testing of the proposed unrolled network, respectively.

### 3.2 Data Pre-Processing

Low resolution gradient echo data were also acquired for coil sensitivity estimation using ESPIRiT[51]. 1D Nyquist ghost correction was applied using a constant gradient-delay and linear phase-error estimated from an EPI calibration scan with phase-encoding gradient turned off. Since cEPI uses variable ESP and ramp-sampling, re-gridding was performed along $k_y$ line-by-line. The central low-resolution k-space data (i.e., matrix size 128x128) of each BUDA shot is reconstructed using SENSE[52]. Cubic interpolation was applied to the low-resolution images of BUDA pairs to create images with the same size as the high-resolution BUDA-cEPI (i.e., matrix size 300x300). These interpolated images were used to estimate the field map via FSL TOP-UP for each diffusion encoding direction, capturing both susceptibility and eddy current effects[53]. This map is referred to as $\Delta\omega_0$ (in unit of radian) as described in Eqs. (1) and (2).

### 3.3 BUDA-cEPI RUN-UP

Unlike the original RUN-UP, BUDA-cEPI RUN-UP is implemented BUDA-cEPI operators ($A_\uparrow$ and $A_\downarrow$) and virtual coil concept to jointly reconstruct the blip-up and blip-down images from the data acquired with the parallel and partial Fourier BUDA-cEPI sequence (i.e., $G_\uparrow$ and $G_\downarrow$) which aim to minimize the following objective function:

$$\frac{1}{2}\min_{u_\uparrow, u_\downarrow} \left\| \begin{bmatrix} A_\uparrow & 0 \\ 0 & A_\downarrow \end{bmatrix} \begin{bmatrix} u_\uparrow \\ u_\downarrow \end{bmatrix} - \begin{bmatrix} G_\uparrow \\ G_\downarrow \end{bmatrix} \right\|_2^2 + R(u_\uparrow, u_\downarrow). \quad (8)$$

In particular, the blip-up and blip-down images are updated using the following equations

$$\begin{aligned} u_{t,\uparrow} &= u_{t-1,\uparrow} - \tau\left(A_\uparrow^H A_\uparrow u_{t-1,\uparrow} - A_\uparrow^H G_\uparrow\right) \\ u_{t,\downarrow} &= u_{t-1,\downarrow} - \tau\left(A_\downarrow^H A_\downarrow u_{t-1,\downarrow} - A_\downarrow^H G_\downarrow\right) \end{aligned} \quad (9)$$

$$\begin{aligned} &\text{Option 1:} & \{u_{t+1,\uparrow}, u_{t+1,\downarrow}\} &= R(u_{t,\uparrow}, u_{t,\downarrow}) \\ &\text{Option 2 (virtual coil):} & \{u_{t+1,\uparrow}, u_{t+1,\uparrow}^*, u_{t+1,\downarrow}, u_{t+1,\downarrow}^*\} &= R(u_{t,\uparrow}, u_{t,\uparrow}^*, u_{t,\downarrow}, u_{t,\downarrow}^*) \\ &\text{Option 3 (virtual coil + b0 images):} & \{u_{t+1,\uparrow}, u_{t+1,\uparrow}^*, u_{t+1,\downarrow}, u_{t+1,\downarrow}^*\} &= R(u_{t,\uparrow}, u_{t,\uparrow}^*, u_{t,\downarrow}, u_{t,\downarrow}^*, b_{0,\uparrow}, b_{0,\downarrow}). \end{aligned} \quad (10)$$

$A^H$ is the adjoint of $A$. $\tau$ is the step size which was manually selected ($\tau = 0.9$). '\*' denotes complex conjugate transpose which is referred to as virtual conjugate coil data [13]. $R(\cdot)$ is a regularization term that is modeled using U-Nets[54].

The proposed model architecture has 3 processing blocks (T = 3 in Fig. 2) which correspond to 6 gradient updates, 3 U-Nets in the image-space, and 3 U-Nets in the k-space. Three options were investigated in this study.

- In option 1, the inputs for U-Nets require only blip-up and blip-down images.
- In option 2, the inputs for U-Nets require blip-up and blip-down images and their virtual coil images. This option is motivated by the limitations of the U-Net architecture to match the S-LORAKS results without making the opposite side of k-space easier to access. After the U-Nets, virtual coil images are collapsed to actual coil images by re-arranging such that $0.5u+0.5(u^*)^*$.
- In option 3, it is similar to option 2, except the pre-computed 10 NEX b0 images obtained by S-LORAKS were added as extra input channels, while these channels were collapsed for the output. This is basically one of the same principles that was use on autocalibrated structured low-rank EPI ghost correction[55], which itself was motivated by multi-contrast reconstruction[56].

To allow different regularization functions for different processing blocks and spaces, the six U-Nets do not share their weights, resulting in the total number of trainable parameters of 12,708,984. Each U-Net consists of convolutional layer with 3×3 kernel size, filter of 64, depth of 3, and dropout of 0.05. We implemented the proposed model in Tensorflow[57] and trained it by minimizing the normalized-root-mean-squared-error (NRMSE) loss between the reconstructed and ground truth blip-up/blip-down images using the Adam optimizer[58] with a learning rate of $1\times10^{-4}$ and batch size of two, running on a 32 GB NVIDIA Quadro GV100 graphics processing unit (GPU). The ground-truth data were prepared using S-LORAKS (20 inner and 15 outer iterations, rank = 80, $\lambda$ = 0.05, and Fourier radius = 3). 5,120 and 1,280 slices from 8 volunteers (whole-brain coverage) were used as the training and validation data, respectively. 800 slices from the 9th volunteer were used for testing the trained model.

**3.4 Experiments**

We performed three experiments to assess the performance of the BUDA operators, S-LORAKS constrained reconstruction, and BUDA-cEPI RUN-UP.

*3.4.1. Simulated BUDA-cEPI with S-LORAKS*

To characterize the performance of S-LORAKS in reconstructing BUDA-cEPI data with a combined RO- & PE-pF sampling, BUDA-cEPI data at 6/8 RO- & PE partial Fourier were simulated from an acquired BUDA-EPI data with PE-only pF (*Data-I*), using a circular sampling mask and k-space cropping in the RO direction (top right in Fig.3). The simulated BUDA-cEPI data were reconstructed using conventional SENSE and S-LORAKS, while the BUDA-EPI data were reconstructed using S-LORAKS and referred to as reference.

*3.4.2. BUDA-cEPI with S-LORAKS*

To compare the geometric accuracy and sharpness when using conventional SENSE versus BUDA framework (Eq. 3) with S-LORAKS (Eq. 4), high resolution BUDA-cEPI with highly accelerated parallel and partial Fourier acquisitions (Data-II) was used.

*3.4.3 BUDA-cEPI RUN-UP*

To compare the reconstruction quality achieved when using conventional SENSE, S-LORAKS (Eq. 4), and BUDA-cEPI RUN-UP (Eq. 6), *Data-II* and *Data-III were* used for training and testing, respectively. To evaluate the robustness and generalizability of the proposed reconstruction, leave-one-subject-out test was performed four times - data from eight and one subjects were used for training, and testing, respectively. To reduce the processing time for the entire reconstruction pipeline, a rapid off-resonance map estimation was also developed using an end-to-end 3D U-Net with 103,668,041 trainable parameters (convolutional layer with 3×3×3 kernel size, filter of 64, depth of 2, and dropout of 0.05). Note that *Data-II* were used for network training of this off-resonance map estimation U-Net - inputs were pair of low-resolution blip-up and -down cEPI images obtained by SENSE, and ground truths were field maps estimated by FSL TOP-UP[10]. Note that NRMSE was simultaneously computed for all slices. Structural similarity index measure (SSIM) and peak signal-to-noise ratio (PSNR) were computed for each slice. Mean and standard deviation (SD) of SSIM and PSNR were reported.

## 4. RESULTS

**4.1 Simulated BUDA-cEPI with S-LORAKS**

Fig. 3 shows that simulated individual blip-up and -down BUDA-cEPI acquisition with conventional SENSE reconstruction resulted in the inability to recover the missing partial Fourier data (3b) – with the resulting image appearing slightly blurry (3e and 3h) compared to the reference (3d and 3g). Moreover, background noise is highly visible. In contrast, the joint reconstruction across the blip-up and -down data that incorporates off-resonance effect and S-LORAKS constraint improves the recovery of the missing k-space data (3c) – where the overall quality of the reconstructed image (3f and 3i) is mostly identical to the reference by visual inspection.

**4.2 BUDA-cEPI with S-LORAKS**

Fig. 4 shows that individual blip-up and -down cEPI with conventional SENSE reconstruction resulted in image blurring at the brain's boundaries (enlarged view images), residual aliasing artifacts, relatively high noise appearance, and geometric distortions (1st row). In contrast, for the joint reconstruction of the blip-up and -down cEPI with BUDA operators and S-LORAKS (2nd row), image boundaries appear sharper, less noise appearance, and no aliasing artifacts are visually detected. Moreover, the geometries of both the blip-up and the blip-down images are well-aligned as shown in overlaid images. However, the reconstruction time of the joint reconstruction is much longer when compared to conventional SENSE (for both polarities), with reconstruction times of 225 and 3.12 seconds, respectively.

**4.3 BUDA-cEPI RUN-UP**

*4.3.1 Non-Diffusion BUDA-cEPI (Data-III)*

The proposed unrolled network was also trained with non-diffusion (b-value 0) images (768/192 slices for training/validation). The same hyperparameters as in the diffusion network training were used (see the methods section). Fig. 5c and 5h shows the error maps in image-space and in k-space for the unrolled reconstruction of the non-diffusion weighted data when the virtual coil concept was not incorporated, resulting in a high RMSE of 15.2%. The reconstructed data in k-space domain appeared inhomogeneous, in which the non-acquired k-space area could not be estimated properly (yellow head arrows in 5g), resulting in a loss of high spatial frequency information, indicated by the yellow head arrows (5c). In contrast, the virtual coil concept enables the unroll network to better take advantage of smooth phase prior to provides improved reconstruction. Low RMSE, with value of 4.1%, was achieved (5e and 5j), where the missing k-space data were effectively recovered (5i).

*4.3.2 Diffusion BUDA-cEPI (Data-III)*

In Fig. 6, unlike the non-diffusion data, for diffusion weighted data, BUDA-cEPI RUN-UP with virtual coil appears insufficient to enable high-fidelity reconstruction. High RMSE values were detected which varied between 29.6%-47.2% among different diffusion directions (6c), with the enlarged view in (6b) highlighting an increased in blurriness of the reconstructed image (yellow headed arrow). In contrast, large reconstruction improvement is achieved when using BUDA-cEPI RUN-UP with both virtual coil and b0 images, where the RMSE values became less than 5.0% (6e). Nonetheless, imperfection in the trained model can leave some small residual artifacts, particularly at the image center – small white spots are slightly visible in the subtraction images (6e).

*4.3.3 Technical Evaluation*

In Table 3, NRMSE values from all models under the conditions of leave-one-subject-out test are lower than 6%. Means and standard deviations of the structural similarity index measure (SSIM) are 0.96±0.01 and 0.97±0.01, respectively. Means and standard deviations of the peak signal-to-noise ratio (PSNR) are 37.94±0.54 and 37.29±0.57, respectively. All three parameters reflect the proposed reconstruction's accuracy, robustness, and generalizability, even with a small training data size (only eight subjects).

*4.3.4 DTI Application*

Fig. 7a shows the estimated eddy current displacement map for a representative slice and diffusion direction obtained using FSL-EDDY. The displacement is larger in areas further away from the iso-center of the scanner, and changes accordingly with diffusion directions as demonstrated in Fig. 7b. The maximum and minimum values of the displacement are -1.5 mm. and +1.9 mm., respectively. These variations are large enough to affect DTI application (7g, and 7h). The geometric inconsistencies are clearly visible after conventional SENSE reconstruction, resulting in blurring on the FA map (7h) and poor alignment of primary eigenvectors on colored FA (7g). The SENSE reconstructed images with FSL-EDDY enable partially managing the geometric distortion, thereby improving the FA maps (7i and 7j). However, the variable ESP at outer k-space, the partial Fourier acquisition, and image domain interpolation during data post-processing cause blurriness in the diffusion images and FA maps. BUDA-cEPI S-LORAKS and BUDA-cEPI RUN-UP reconstructions performed to the same level from visual inspection (7k vs. 7m and 7l vs. 7n), and outperformed conventional SENSE in reducing residual artifacts, and enhancing small details, resulting in improved diffusion images and FA maps.

Fig. 8 demonstrates the capability of BUDA-cEPI S-LORAKS and BUDA-cEPI RUN-UP in recovering imaging details due to partial Fourier acquisition. Recovering the information allows for visualizing more details of fiber orientation distributions in cortical areas, as shown in the enlarged views (8h and 8i).

*4.3.5 Time-efficient Reconstruction Pipeline*

Table 3 demonstrates that BUDA-cEPI RUN-UP significantly improves the reconstruction time (2.54 seconds) which is about 88x faster than BUDA-cEPI S-LORAKS (225.32 seconds). However, the time required for field map estimation using FSL TOP-UP took roughly 12 seconds per slice. The use of 3D U-Net enables a time reduction for the field map estimation task to only 0.05 second per slice, thereby reducing the overall reconstruction time by 80% (3.03 vs. 15.06 seconds). Note that the field differences between 3D U-Net and TOP-UP based field maps were very small (less than 3% of NRMSE). Consequently, the incorporation of 3D U-Net based field maps in BUDA-cEPI RUN-UP did not compromise overall quality of reconstructed images (the results not shown).

## 5. DISCUSSION AND CONCLUSION

In this study, we developed a rapid ML-based reconstruction approach for distortion-free high resolution dMRI with BUDA-cEPI acquisition. A model-based framework that manages for geometric distortions caused by off-resonance effects was unrolled through a tailored artificial neural network with only six gradient updates. The reconstruction was shown to significantly reduce the reconstruction time, while providing high quality results comparable to that of the state-of-the-art technique, S-LORAKS[48].

Among various constrained MRI reconstruction techniques, S-LORAKS[48] was chosen for this work. This approach takes advantage of multiple constraints jointly, such as limited image support, slowly varying phase, multi-channel[24] and/or multi-echo[56] acquisition. Those constraints lead to shift-invariant autoregressive prediction relationships in k-space, inducing low-rankness of the corresponding structured low-rank matrices[48,59,60] Our results also demonstrate that S-LORAKS is well-suite for parallel and partial Fourier BUDA-cEPI acquisition – residual aliasing artifact and partial Fourier effect are effectively suppressed, thereby achieving better SNR and sharpness. However, one of the challenges associated with implementing the LORAKS reconstruction is that the matrix $P_{s\uparrow\downarrow}(u_\uparrow, u_\downarrow)$ is many times larger than the original images $u_\uparrow$ and $u_\downarrow$. The step in building such matrix can be relatively slow and computationally intensive. It has recently been observed[29] that the convolutional structure of this kind of matrix allows computations involving $P(\cdot)$ to be performed using simple convolution operations (which can also be implemented efficiently using the FFT), without the need for explicitly forming the large-size LORAKS matrix. However, it requires approximately 55 seconds of processing time for data with a matrix size of 255x255, which may not be fast enough for clinical practice.

In this work, the use of an unrolled supervised learning algorithm is chosen to accelerate the reconstruction process, where such a network was tailored both in term of its unrolled structure and the incorporation of virtual coils to enable it to perform well for BUDA-cEPI. This approach is inspired by classic variational optimization methods and iterate between data-consistency enforcement and deep learning model that acts as a regularizer[35-37]. It allows flexibility in trading off between the number of iterations (data consistency blocks) and trainable parameters. Recently, Hu Y. et al[38] reported that RUN-UP enabled nearly real-time reconstruction and improved image quality for brain and breast DWI applications compared to images obtained by conventional reconstruction. Their network unrolled 6 iterations of FISTA with a total of 2,396,454 parameters. Aggarwal H. et al[37] developed MoDL-MUSSELS that also implemented standard SENSE for data consistency. They unrolled 5 outer and 5 inner iterations of the IRLS algorithm. In this study, we implemented 6 gradient updates (6 U-Nets, 3 for image-space and 3 for k-space) with trainable parameters of 12,708,984. Typically, the employment of an extensive set of trainable parameters has been observed to substantially enhance the attainment of precise outcomes in the context of intricate tasks. Nevertheless, this practice is concomitant with inherent perils, notably overfitting and the occurrence of vanishing gradients, both of which can lead to the inadequate training of neural networks. In such case, some hyper-parameters may be carefully fine-tuned. The selection of proper dropout rate[61] is often mentioned.

The proposed method, which incorporates the virtual coil (VC) data, improves results as demonstrated in Figs. 5d and 5i. The utilization of VC technique represents a highly efficacious strategy for augmenting the performance of parallel MRI[62], with particular relevance in scenarios involving echo planar imaging (EPI) employing partial Fourier acquisition. VC achieves the generation of virtual coils through the assimilation of conjugate symmetric k-space signals derived from physical coils, thus augmenting the available information to address gaps in k-space data, a feature particularly advantageous in conjunction with partial Fourier acquisition. In essence, the implementation of VC consistently ensures image quality on par with or superior to that of images reconstructed without VC. Recently, Cho J. et al[63] presented evidence of a network that incorporates convolutional neural network (CNN) denoisers in both k-space and image-space domains, harnessing the potential of virtual coils to enhance the conditioning of image reconstruction. Furthermore, our findings (Fig. 6) indicate that further adding non-diffusion images as an additional channel can enhance the network's performance. Previous studies have also shown that including supplementary contrasts, apart from diffusion-weighted images, in the input data for the learning algorithm aids in delineating anatomical boundaries with preventing blurring artifacts in the outputs[64,65].

As shown in Table 3, RUN-UP BUDA is robust and generalizable across subjects as demonstrated through NRMSE, SSIM, and PSNR. For model accuracy, this was reflected through the DTI application where the results obtained by BUDA-cEPI RUN-UP and BUDA-cEPI S-LORAKS appeared comparable (Figs. 7 and 8). Even though we have shown that our BUDA-cEPI RUN-UP can work well and is robust for the same protocol across subjects, the robustness of using this reconstruction model could decrease when applied to acquisition with protocols that has significantly different resolution and/or noise distribution. This is a general issue that has been discussed in detail in recent works[66,67]. Fabian Z. et al[66] introduced a physics-based data augmentation pipeline for accelerated MR imaging.

This strategy showed the robustness against overfitting and shifts in the test distribution. Knoll F. et al[67] demonstrated that by increasing the heterogeneity of the training data set, trained networks can be obtained that generalize toward wide-range acquisition settings, including contrast, SNR, and particular k-space sampling patterns. Their study also provides an outlook for the potential of transfer learning to fine-tuning of our network to a particular target application using only a small number of training cases.

The proposed BUDA-cEPI RUN-UP integrates off-resonance effect through time segmentation strategy[47]. In addition to the number of time segmentations, the number of coil and the resolution of acquired data are proportionally relative to the reconstruction time. Our technique took longer (i.e., 3.03 seconds) than RUN-UP[38] and MoDL-MUSSELS[37] (i.e., 0.1, and 0.16 seconds, respectively) that off-resonance was not considered. It is worth noting that the extension of the input channel with virtual coil data had only a very slight impact on the reconstruction time, as this step is performed after all coil data have been combined. An advanced coil compression and/or coil sketching techniques[68] could further reduce coil channels, which may further improve the speed of BUDA-cEPI RUN-UP.

While machine learning (ML) reconstructions have proven beneficial in reducing noise[69], they might compromise spatial resolution[70]. Future research will delve into using high SNR ground truth data sourced from multiple averaged captures to train the network in reconstructing and denoising single average captures. In diffusion data, every reconstructed image will display varied phase variations between shots, necessitating the use of real-valued averages to create accurate ground truth devoid of magnitude noise bias[71]. Additionally, because an image reconstructed from a single average will have a distinct background phase relative to the ground-truth data, we'll have to modify the training cost function. The background phase from the single-average reconstruction will have to be eliminated prior to its comparison with the ground truth.

In conclusion, we developed a new reconstruction pipeline, called BUDA-cEPI RUN-UP, for parallel and partial Fourier BUDA-cEPI acquisition. This proposed technique uses a deep-learning architecture, combining an MR-physic model (BUDA-cEPI operators) and U-Nets in both k-space and image space as trainable priors, with virtual coil concept also incorporated. Such technique was shown to reduce the reconstruction time by ~88x when compared to the state-of-the-art technique, while preserving imaging details as demonstrated through DTI application.

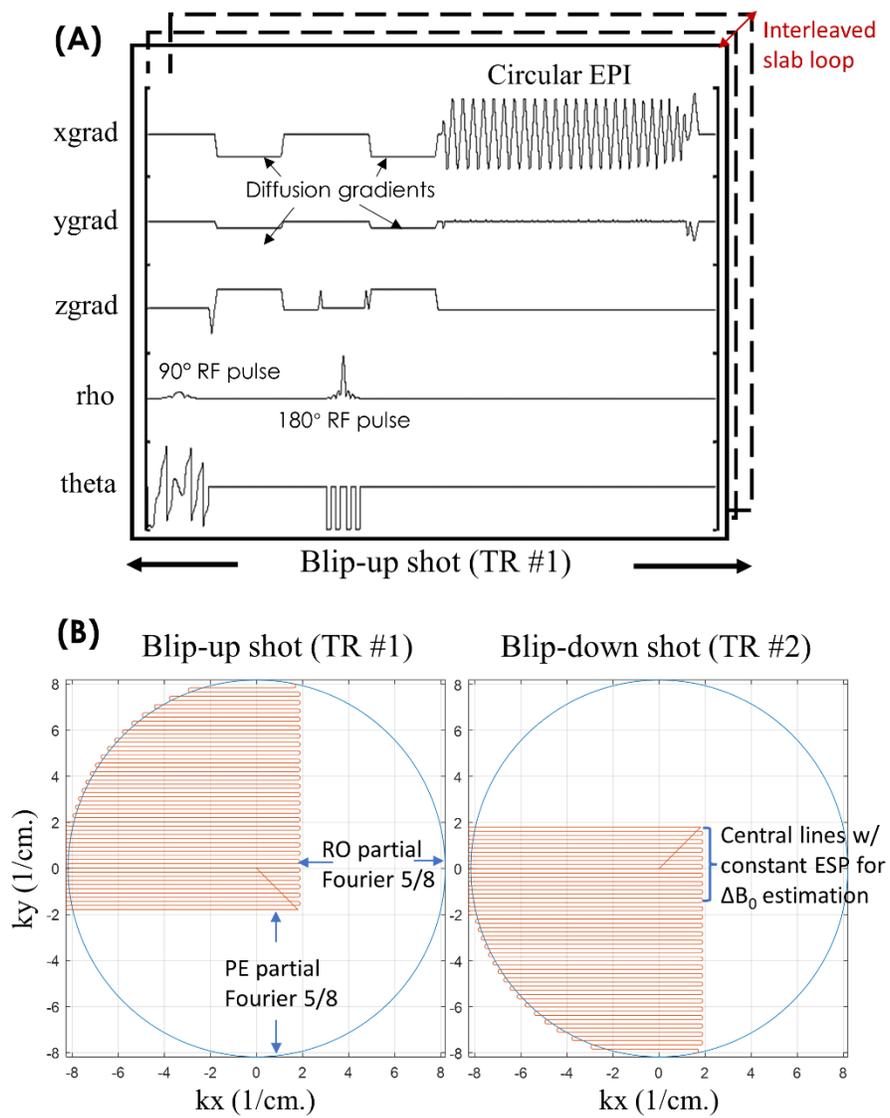

**Fig. 1.** (A) The sequence diagram of the BUDA-cEPI sequence. (B) The trajectory of the blip-up and blip-down cEPI with readout and phase-encoding partial Fourier acquisition.

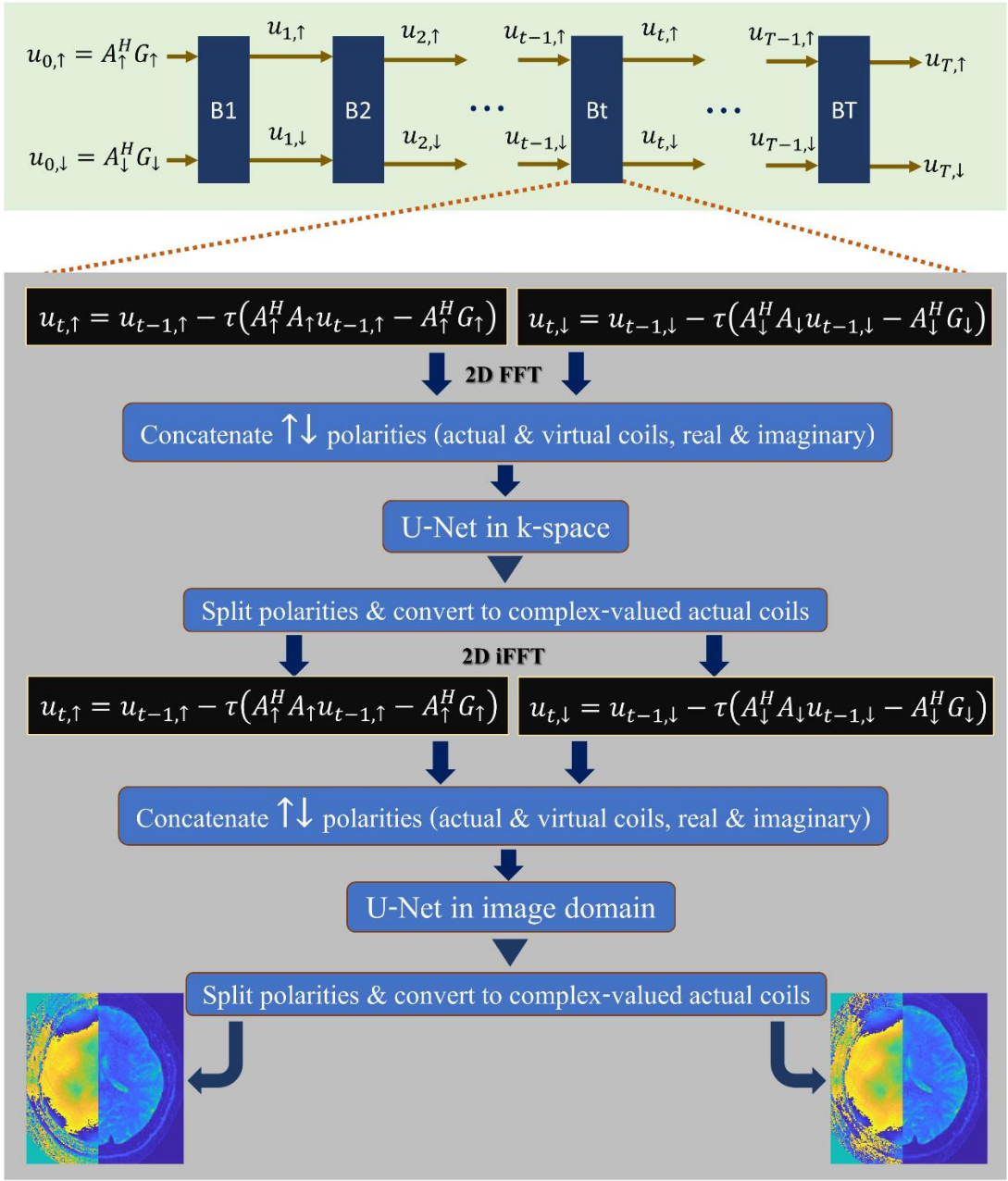

**Fig. 2.** The proposed unrolled network reconstruction for BUDA-cEPI (BUDA-cEPI RUN-UP)

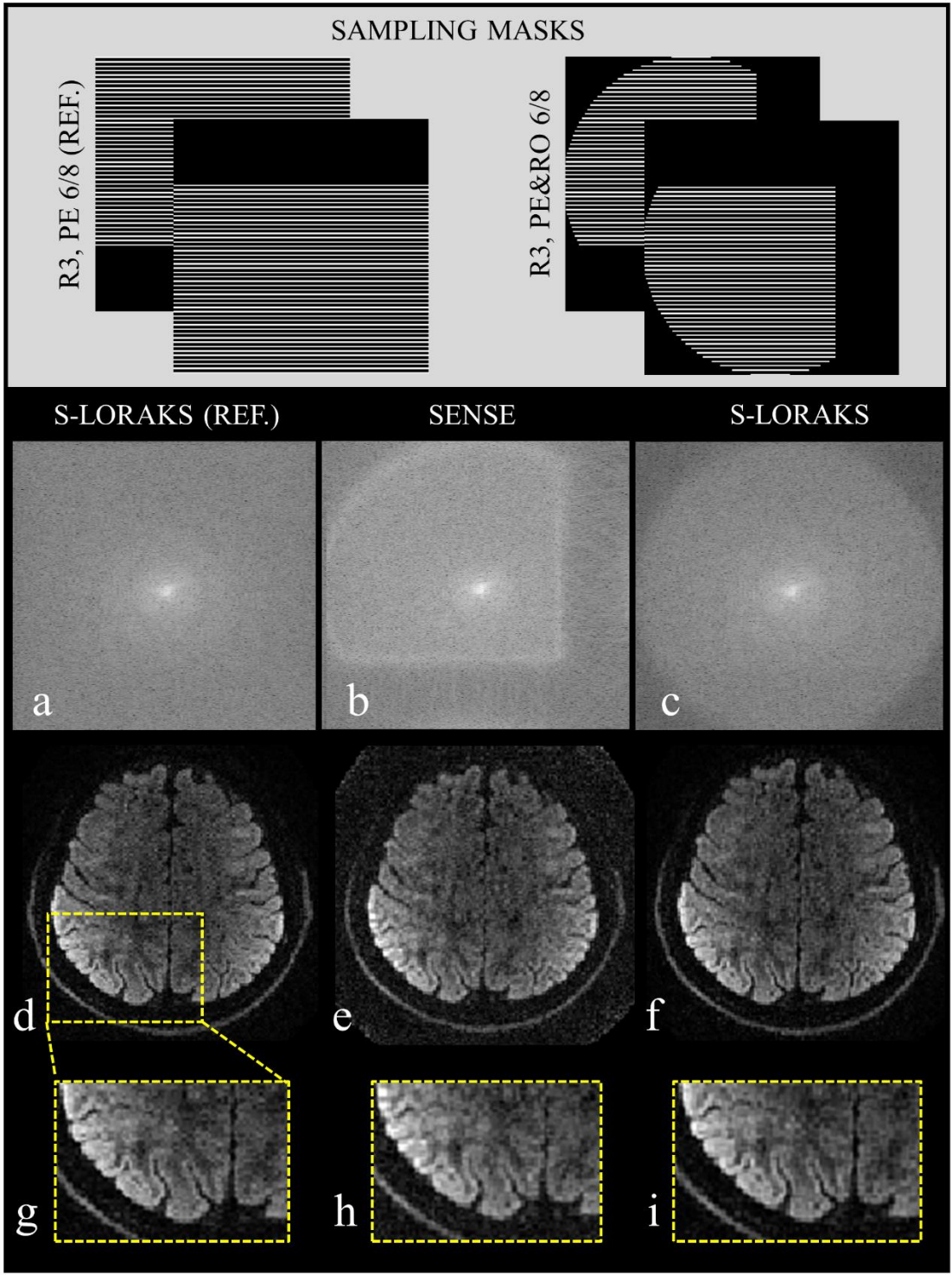

**Fig. 3.** (a-c) reconstructed data distribution in k-space. (d) Images corresponding to Cartesian sampling pattern reconstructed by S-LORAKS. (e,f) Images corresponding to the circular sampling pattern reconstructed by SENSE and S-LORAKS, respectively. (g-i) Enlarged views corresponding to d-f displayed to show fine details.

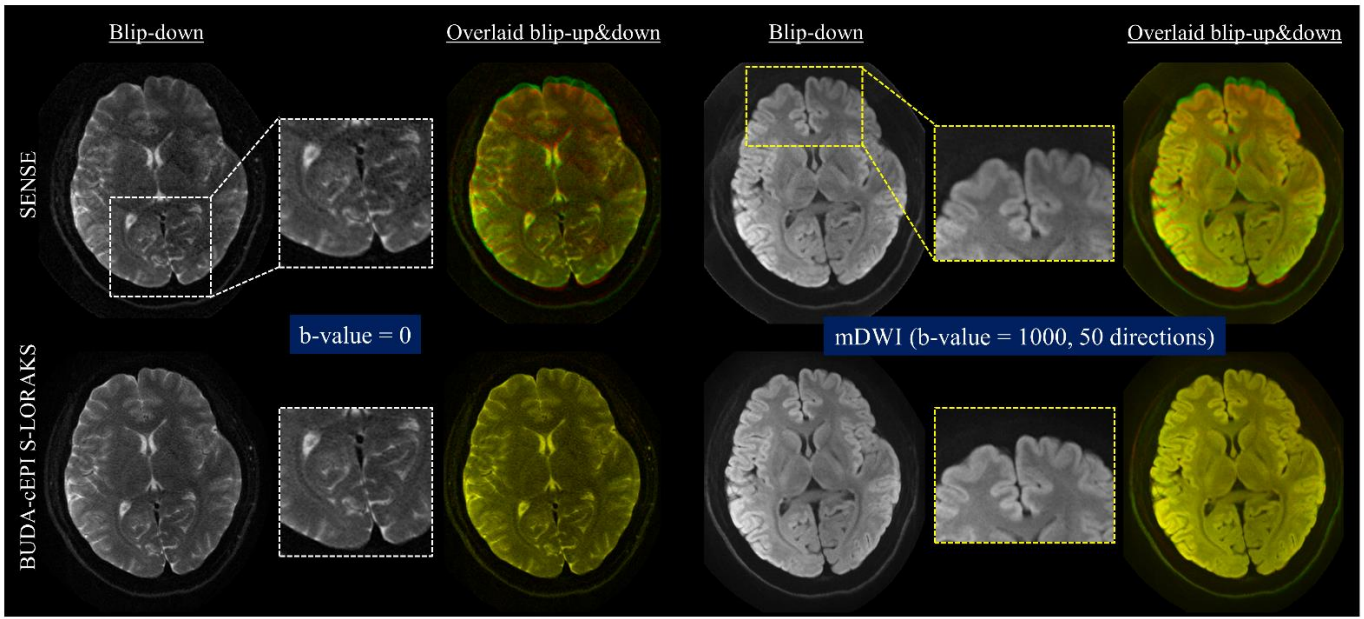

**Fig. 4.** (1st row) Images obtained by standard SENSE. (2nd row) Images obtained by BUDA S-LORAKS. Enlarged views in white and yellow boxes highlight the sharpness at image boundaries. Overlay of the EPI blip-up (green channel) and EPI blip-down (red channel) displayed to demonstrate the geometry alignment.

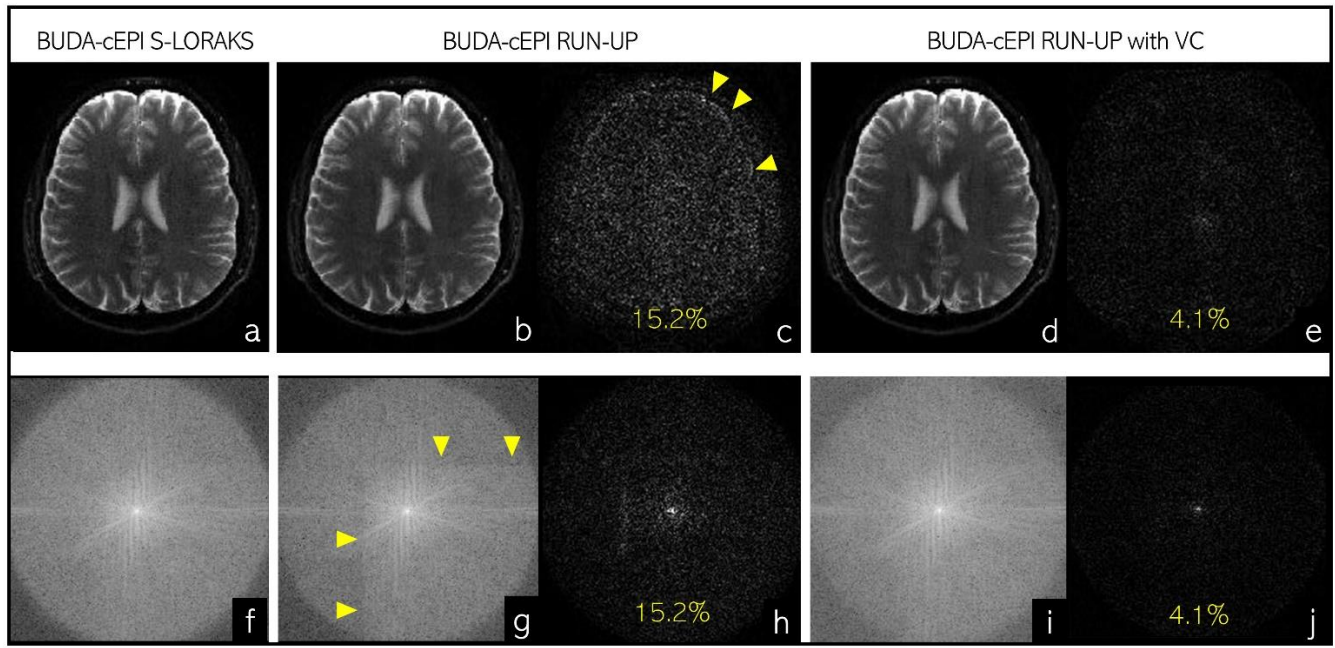

**Fig. 5.** All reconstructed images are from the same b-value 0 data acquired using BUDA-cEPI. (a) image obtained by S-LORAKS. (b) image obtained by unrolled KI-Net. (c) the difference between a and b. (d) image obtained by unrolled KI-Net with virtual coil data. (e) the difference between a and d. (f) k-space data corresponding to a. (g) k-space data corresponding to b. (h) the difference between f and g. (i) k-space data corresponding to a. (j) the difference between f and i. The superimposed numbers on c, e, h, and j are %RMSE, in which BUDA S-LORAKS (a) was used as a reference.

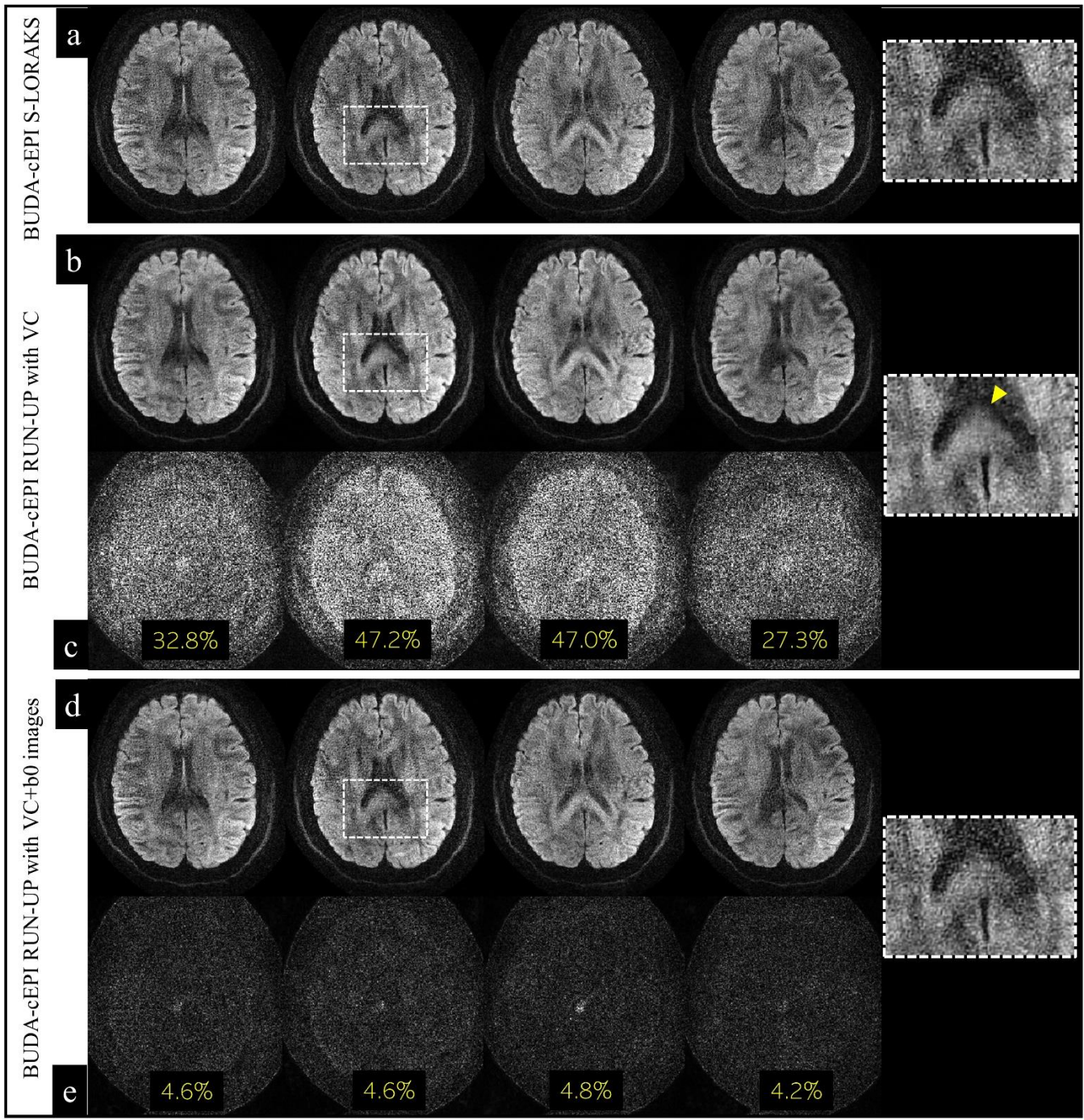

**Fig. 6.** Reconstructed images in rows (a, b, and d) are from the same data set acquired using BUDA-cEPI b-value 1000 s/mm$^2$ at four different diffusion directions. (a) images obtained by S-LORAKS and referred to as reference for computing NRMSE. (b) image obtained by unrolled KI-Net with virtual coil. (c) the difference between a and b. (d) image obtained by unrolled KI-Net with virtual coil and b-value 0 channels so-call RUN-UP BUDA. (e) the difference between a and d. The superimposed numbers on c and e are %RMSE.

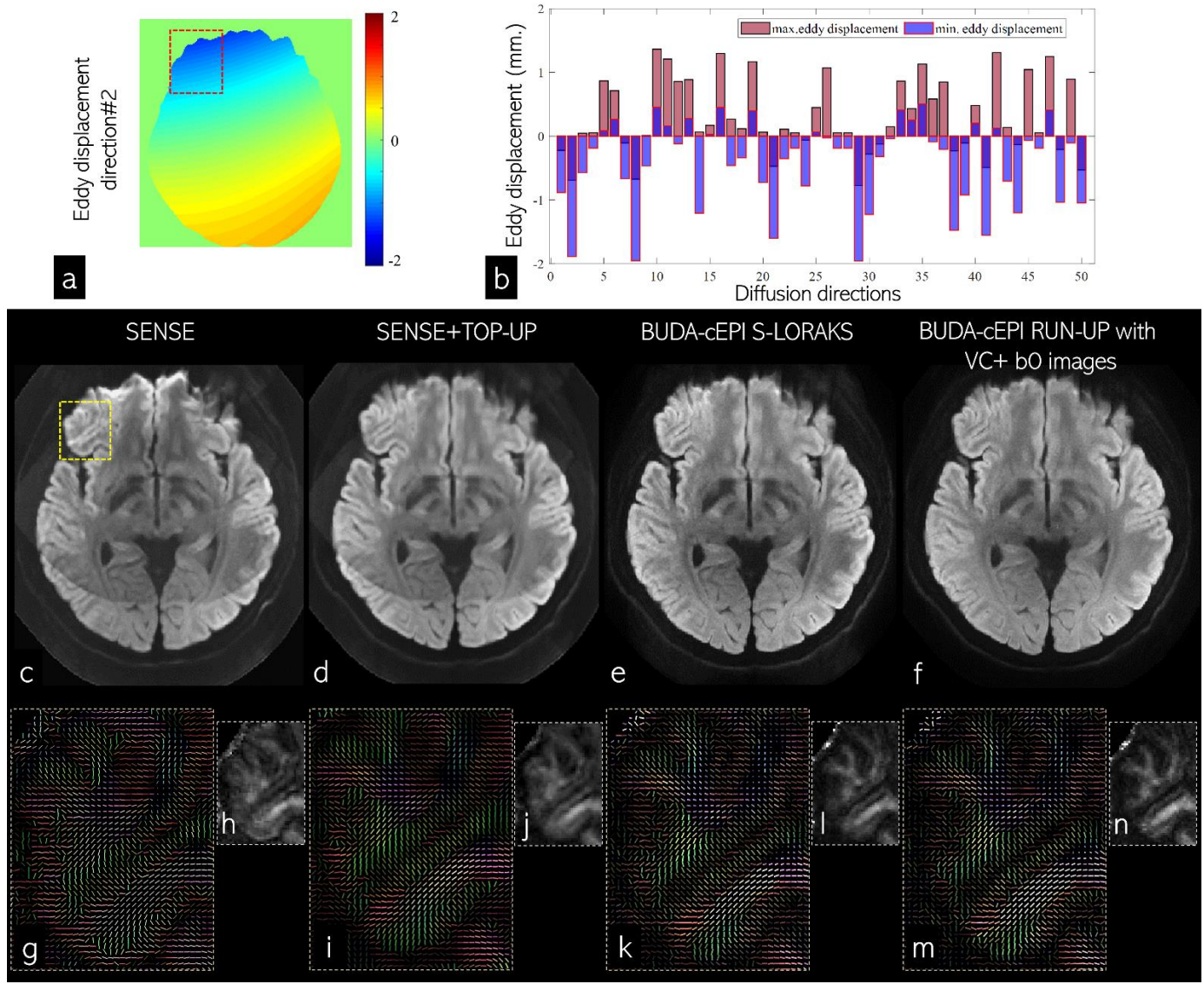

**Fig. 7.** (**a**) one representative eddy displacement obtained by FSL-EDDY. (**b**) bar plots of maximum and minimum eddy displacement inside the red box area in (a) across 50 diffusion directions. (**c-f**) mean diffusion images. (**g, i, k, and m**) the primary eigenvectors at yellow box area corresponding to each reconstruction technique were color-encoded (red: left-right, green: anterior-posterior, blue: superior-inferior). (**h, j, l, and n**) FA maps without directional information at yellow box area corresponding to each reconstruction technique.

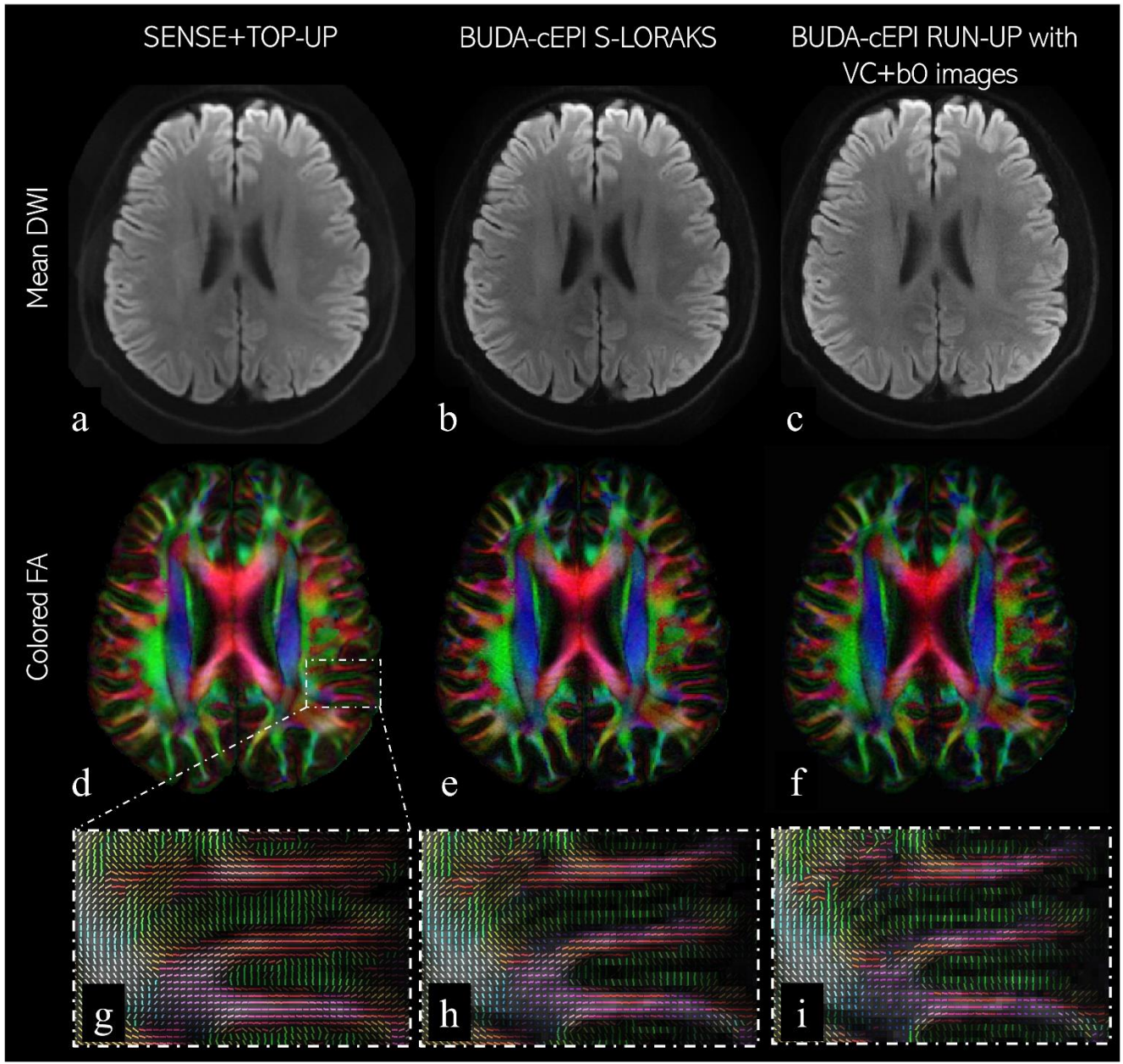

**Fig. 8.** (a-c) mean diffusion images. (d-f) colored FA maps (red: left-right, green: anterior-posterior, blue: superior-inferior) corresponding to diffusion images in a-c, respectively. (g-h) enlarged views of the primary eigenvectors were color-encoded and overlaid on FA maps. The green fibers in (g) show overlapping cortex area across two gyri.

**Table 1.** Imaging sequences and parameters.

| Parameters | Data-I | Data-II | Data-III |
| --- | --- | --- | --- |
| Sequence | BUDA-EPI | BUDA-cEPI | BUDA-cEPI |
| Resolution [mm] | 1.25x1.25x2.00 | 0.73x0.73x5.00 | 0.73x0.73x5.00 |
| Repetition time (TR)[msec.] | 2800 | 5000 | 5000 |
| Echo Time (TE) [msec.] | 77 | 55 | 55 |
| Field of View (FOV) [mm.] | 220x220 | 220x220 | 220x220 |
| Matrix size | 176x176 | 300x300 | 300x300 |
| Number of Slice | 57 | 16 | 16 |
| Echo Spacing (ESP) [msec.] | 1.11 | variable ESP, from 1.09 to 0.67 | variable ESP, from 1.09 to 0.67 |
| Partial Fourier | 6/8 | 5/8 | 5/8 |
| SENSE | 3 | 4 | 4 |
| Number of Excitation (NEX) | 1 | 1 | 3 |
| Scan time [sec.] | 210 | 300 | 900 |
| Number of volunteers | 1 | 8 | 1 |
| Number of b-value 0 (sec./mm.$^2$) | 10 | 10 | 10 |
| Number of b-value 1000 (sec./mm.$^2$) | 64 | 50 | 50 |

**Table 2.** The results of leave-one-subject-out test. Single value of normalized root-mean-squares-error (NRMSE) was reported. It was computed simultaneously for all slices and diffusion directions. Structural similarity index measure (SSIM) and peak signal-to-noise ratio (PSNR) were computed slice-by-slice. Mean and SD values of SSIM and PSNR across all slices and diffusion directions were reported.

| TRAIN (8 subjects) | TEST (1 subject) | %NRMSE | SSIM (Mean±SD) | PNSR (Mean±SD) |
|---|---|---|---|---|
| exclude subject 1 | subject 1 | 5.57 | 0.96±0.01 | 37.66±0.45 |
| exclude subject 2 | subject 2 | 5.32 | 0.97±0.01 | 37.29±0.57 |
| exclude subject 3 | subject 3 | 5.26 | 0.96±0.01 | 37.94±0.54 |
| exclude subject 4 | subject 4 | 5.35 | 0.97±0.01 | 37.57±0.42 |

**Table 3.** Processing times per slice (in second) for four different reconstruction pipelines.

| | Field map estimation (matrix 128x128) | | | Reconstruction | Total |
|---|---|---|---|---|---|
| | SENSE | TOP-UP | 3D U-Net | | |
| BUDA-SLORAKS-I | 0.44 | 12.08 | - | 225.32 | **237.84** |
| BUDA-SLORAKS-II | 0.44 | - | 0.05 | 225.32 | **225.81** |
| RUN-UP BUDA-I | 0.44 | 12.08 | - | 2.54 | **15.06** |
| RUN-UP BUDA-II | 0.44 | - | 0.05 | 2.54 | **3.03** |